\documentclass[12pt]{article}
\usepackage{amsmath}
\usepackage{graphicx}
\usepackage{amsfonts}
\usepackage{amssymb}

\begin{document}
\input epsf

\def\half{{\textstyle{1\over2}}}
\let\a=\alpha \let\b=\beta \let\g=\gamma \let\d=\delta \let\e=\epsilon
\let\z=\zeta \let\h=\eta \let\th=\theta \let\i=\iota \let\k=\kappa
\let\l=\lambda \let\m=\mu \let\n=\nu \let\x=\xi \let\p=\pi \let\r=\rho
\let\s=\sigma \let\t=\tau \let\u=\upsilon \let\f=\phi \let\c=\chi \let\y=\psi
\let\vp=\varphi \let\vep=\varepsilon
      \let\G=\Gamma \let\D=\Delta \let\Th=\Theta \let\L=\Lambda
\let\X=\Xi \let\P=\Pi \let\S=\Sigma \let\U=\Upsilon \let\Y=\Psi
\let\C=\Chi
\let\la=\label \let\ci=\cite \let\re=\ref
\let\se=\section \let\sse=\subsection \let\ssse=\subsubsection
\def\nn{\nonumber} \def\bd{\begin{document}} \def\ed{\end{document}}
\def\ds{\documentstyle} \let\fr=\frac \let\bl=\bigl \let\br=\bigr
\let\Br=\Bigr \let\Bl=\Bigl
\let\bm=\bibitem
\let\na=\nabla
\def\tU{{\widetilde U}}
\let\pa=\partial \let\ov=\overline
\def\ie{{\it i.e.\ }}
\def \be {\begin{equation}}
\def \ee {\end{equation}}
\def\ba{\begin{array}}
\def\ea{\end{array}}
\def\ft#1#2{{\textstyle{{\scriptstyle #1}\over {\scriptstyle #2}}}}
\def\fft#1#2{{#1 \over #2}}
\def\F#1#2{{ F_{#1}^{(#2)} }}
\def\cF#1#2{{ {\cal F}_{#1}^{(#2)} }}

\def\R{{\bf R}}
\def\sst#1{{\scriptscriptstyle #1}}
\def\oneone{\rlap 1\mkern4mu{\rm l}}
\def\e7{E_{7(+7)}}
\def\td{\tilde}
\def\wtd{\widetilde}
\def\im{{\rm i}}
\def\bog{Bogomol'nyi\ }

%%%%%%%%%%%%%%%%%%%%%%%% Greek Letters %%%%%%%%%%%%%%%%%%%%%%%%%%%%%%%%%%

\def\l {\lambda}
\def\a {\alpha}
\def\ap {\alpha'}
\def\b {\beta}
\def\g {\gamma}
\def\G {\Gamma}
\def\d {\delta}
\def\s {\sigma}
\def\e {\epsilon}
\def\vt {\vartheta}
\def\vp {\varphi}
\def\T {\Theta}

\renewcommand{\O}{\Omega}
\renewcommand{\L}{\Lambda}
\renewcommand{\t}{\theta}

%%%%%%%%%%%%%%%%%% Calligraphic Letters %%%%%%%%%%%%%%%%%%%%%%%%%%%%%%%%%

\def \rt {{\rm t}}
\def \ci{\cite}
\def \YY {{\rm Y}}
\def \lra {\leftrightarrow}
\def \const {{|rm const}}

\def\be{\begin{equation}}
\def\ee{\end{equation}}
\def\ba{\begin{eqnarray}}
\def\ea{\end{eqnarray}}

\def\dg{\dagger}
\def\a{\alpha}
\def\b{\beta}
\def\e{\varepsilon}
\def\p{\phi}
\def\ap{\alpha^\prime}
\def\I{{\cal I}}

\def\R{{\bf R}}
\def\Z{{\bf Z}}
\def\C{{\bf C}}
\def\P{{\bf P}}
\def\xb{{\bar X}}
\def\Tr{{\rm  Tr}}
\def\tr{{\rm  tr}}
\def \del{\partial}
\def \a {\alpha}
\def \aa {{\a'}}
\def\g{\gamma}
\def\s{\sigma}
\def\z{\zeta}
\def\zi{\zeta_1}
\def\zii{\zeta_2}
\def\ov{\over}
\def\la{\label}
\def\I{{\cal I}}
\def\J{{\mathcal J}}
\def \ok {{1\ov \k}}
\def\LL{{\mathcal L }}
\def \jL {{J}}
\def \om {\omega}
\def \cL {{\mathcal L}} \def \cH {{\mathcal H}}
\def\E{{\mathcal E}}
\def\w{\omega}
\def\b{\beta}
\def\l{\lambda}
\def\eps{\epsilon}
\def\vep{\varepsilon}
\def \De {{\mathcal D}}

%%%%%%%%%%%%%%%%%%%%%%%
\def \adss{$AdS_5 \times S^5$\ }

\def \r { \rho}
\def \sql {\sqrt{\lambda} }
\def \t {\theta}
\def \p {\phi}
\def \vp {\varphi}
\def \Om {\Omega}
\def \ads {{$AdS_5$}}
\def \ov {\over}
\def \s{\sigma}
\def \pa{\partial}
\def \ta{\tau}
\def \sh {\sinh}
\def \ha {{1 \over 2}}

\def \la{\label}
\def  \Jt {  {J}_{\rm tot}    }

\def \k {\kappa}
\def\foot{\footnote}
\def \four{{\textstyle {1\ov 4}}}
 \def \third { \textstyle {1\ov 3}}
\def\det{\hbox{det}}
\def \ci {\cite}

\def \foot {\footnote}
\def \bi{\bibitem}
\def \tr {{\rm tr}}
\def \ha {{1 \over 2}}
\def \tid {\tilde}
\def \vv {{\rm v}}
%%%%%%%%%%%%%%%%%%%%%%%%%%%%%%%%%%%%%%%%%
\def \tl {{\tilde \l}}
\def \XX {{\rm X}}
\def \ta {{\tilde \a}}
\def \fo { {1\ov 4}}
\def \ep {\epsilon}
\def \inti {{\int^{2\pi}_0 {d \sigma \ov 2 \pi}}}

\def \d {\partial}
\def \K {{\rm S}}
\def \el {\ell}
\def \Tr {{\rm Tr}}
\def \P {\Phi}
\def \l  {\lambda}
\def \tl {{\tilde \l}}
\def \bl {{\tilde \l}}
\def \const {{\rm const}}
%%%%%%%%%%%%%%%%%%%%%%%%%%%%%%%%%%%%%%%%%%%%%%%%%%%%%%%%%%%%%%
\def \V {v}

\def \bv {v^*}
\def \vv {{\rm v}}
\def \LL {{\mathcal L}}
\newcommand{\PV}[1]{P_{\!\!_{V_{#1}}}}

\def \bL {\ell}
\def \M {{\mathcal M}}
\def \N {{\mathcal N}}
\def \S {{\rm S}}
\def \vn {\vec n}
\def \tl {\td \l}
\def \td {\tilde}
\def \Prod {\Pi}\def \O {{\mathcal O}}
\def \Q {{\rm  Q}}
\def \D {\Delta}
\def \N {{\mathcal N}}

\def\ve{\varepsilon}
\def\vf{\varphi}
\def\F{\Phi}
\def\wg{\wedge}

\def\ve{\varepsilon}
\def\vf{\varphi}
\def\F{\Phi}
\def\wg{\wedge}

\newcommand{\auth}{AUTHORS}
\def\thb{\bar{\theta}}
\def\Thb{\bar{\Theta}}
\def\barp{\bar{p}}
\def\barq{\bar{q}}
\def\barc{\bar{c}}
\def\bard{\bar{d}}
\def\e{\epsilon}

\def\th{\theta}
\def\Th{\Theta}
\def\vth{\vartheta}
\def\btheta{{\bar\theta}}
\def\ttheta{{{\tilde\theta}}}
\def\bttheta{{{\bar\ttheta}}}
\def\vth{\vartheta}

\def\ra{\rightarrow}
\def\N{{\cal N}}
\def\F{{\cal F}}
\def\uM{\underline{M}}
\def\uN{\underline{N}}
\def\uP{\underline{P}}
\def\cc{\circ}
\def\eqv{\equiv}

\def\ni{\noindent}

\def\Ep{E^{{}^{(+)}}}
\def\Em{E^{{}^{(-)}}}

\def\Mp{M^{{}^{(+)}}}
\def\Mm{M^{{}^{(-)}}}

\def\xb{\bar{x}}
\def\xib{\bar{\xi}}
\def\lb{\bar{\l}}
\def\tb{\bar{t}}
\def\vb{\bar{v}}

\def\an{|n|}
\def\xt{\tilde{x}}
\def\pnr{(p_n^{(r)}}
\def\Xd{\dot{X}}
\def\amn{{\a_{-n}}}
\def\At{\tilde{A}}
\def\Bt{\tilde{B}}
\def\ola{\overleftarrow}
\def\ora{\overrightarrow}
\def\at{\tilde{\a}}
\def\st{\star}
\def\qb{\bar{q}}
\def\qt{\tilde{q}}
\def \w {\omega}

\def \adss {$AdS_5 \times S^5\ $}

\def\ss{\sin^2 \g_0}
\def\ov{\over}
\def \ha { { 1 \over 2}}
\def \kk {{\rm k}}
\def \om {\omega}
\def \Om {\Omega}
\def \X {{\rm  X}}
\def \ww { {\rm w} }
\def \JJ {{\cal I}}
\def \bx {{\bar X}}
\def \del {\pa}
\def \rk {\mbox k}

%%%%%%%%%%%%%%%%%%%%%% Miscellaneous  %%%%%%%%%%%%%%%%%%%%%%%%%%%%%%%%%%%

\newcommand{\bea}{\begin{eqnarray}}
\newcommand{\eea}{\end{eqnarray}}

\newcommand{\beqr}{\begin{displaymath}}
\newcommand{\eeqr}{\end{displaymath}}
\newcommand{\beqa}{\begin{eqnarray}}
\newcommand{\eeqa}{\end{eqnarray}}
\newcommand{\beqar}{\begin{eqnarray*}}
\newcommand{\eeqar}{\end{eqnarray*}}
\renewcommand{\k}{\kappa}

\def\lt {\tilde{\lambda}}
\def\rt {\tilde{r}}
\def\rhot {\tilde{\rho}}
\def \rvac{r_\mt{vac}}
\def \S{{\cal S}}

\newcommand{\rf}[1]{(\ref{#1})}
%%%%%%%%%%%%%%%%%%%%%%%%%%%%%%%%%%%%%%%%%%%%
%Arkady
%%%%%%%%%%%%%%%%%%%%%%%%%%%%%%%%%%%%%%%%%%%%%%%'

\def\appendix#1{
  \addtocounter{section}{1}
  \setcounter{equation}{0}
  \renewcommand{\thesection}{\Alph{section}}
  \section*{Appendix \thesection\protect\indent \parbox[t]{11.15cm}
  {#1} }
  \addcontentsline{toc}{section}{Appendix \thesection\ \ \ #1}
  }

\overfullrule=0pt
\parskip=2pt
\parindent=12pt
\headheight=0in \headsep=0in \topmargin=0in
\oddsidemargin=0in
\vspace{ -2cm}
\begin{flushleft}
\hfill hep-th/0408187
\end{flushleft}

\begin{flushleft}
\hfill OHSTPY-HEP-T-04-004
\end{flushleft}
\vspace{-1cm}

\begin{flushleft}
\hfill NSF-KITP-04-102
\end{flushleft}
%\vspace{-2cm}

\bigskip
\thispagestyle{empty}

\vspace{1cm}
\begin{center}
{\Large\bf On one-loop correction to energy\\
\vspace{0.1cm}
 of
 spinning strings in $S^5$}

\vspace{0.2cm}

\vspace{.5cm} {S.A. Frolov$^{a,}$\footnote{Also at Steklov
Mathematical Institute, Moscow. frolovs@sunyit.edu}, I.Y.
Park$^{b,}$\footnote{ipark@pacific.mps.ohio-state.edu } and A.A.
Tseytlin$^{b,}\footnote{Also at Imperial College London
and  Lebedev  Institute, Moscow. tseytlin@mps.ohio-state.edu}$}\\
\vskip 0.3cm

{\em $^{a}$Department of Applied Mathematics,
SUNY Institute of Technology,\\
P.O. Box 3050, Utica, NY 13504-3050, USA\\
\vskip 0.1cm
$^{b}$Department of Physics, The Ohio State University,\\
 Columbus, OH 43210, USA
       }

\end{center}

 \vspace{0.2cm}
 \begin{abstract}
We revisit the computation (hep-th/0306130)   of
1-loop  $AdS_5 \times S^5$
superstring sigma model correction to   energy of  a   closed  circular
string  rotating in
$S^5$.  The string is spinning around  its center of mass with
  two equal  angular momenta $J_2=J_3$
 and its center of mass angular momentum is $J_1$.
  We revise the argument in hep-th/0306130 that the
1-loop correction is suppressed by $1 \ov J$ factor
($J= J_1 + 2 J_2$ is the total $SO(6)$ spin)
 relative to the classical ${ \lambda J_2\ov J^2}$ term in
the energy and use numerical methods to compute the leading 1-loop
coefficient. The corresponding  gauge theory result is  known
(hep-th/0405055) only in  the  $J_1=0$  limit   when the string
solution becomes  unstable  and thus   the 1-loop shift of the energy
formally contains   an imaginary part.
While the  comparison   with gauge
theory may not be well-defined
in this case, our  numerical  string theory  value  of  the 1-loop
coefficient   seems to  disagree with the
gauge theory one. A  plausible  explanation should be  (as in
hep-th/0405001)  in the different  order of limits taken on the gauge
theory and  the string theory sides of the AdS/CFT duality.

\end{abstract}

\newpage

\setcounter{equation}{0}
\setcounter{footnote}{0}
\setcounter{section}{0}

\section{Introduction}
%%%%%%%%%%%%%%%%%%%%%%%%%%%%%%%%%%%%%%%%%%%%%%%%%%
Recently,  there was an interesting progress in
  understanding AdS/CFT duality
  by extending the BMN approach \ci{bmn} to
other sectors of semiclassical \ci{gkp}  string states
(see \ci{rev} for reviews and references).

In general, for a classical rotating closed string solution in
$S^5$ its energy has a regular expansion \ci{ft2,ft3,ft4,afrt,art}
$E_0= J + c_1 {\fr{\l}{J}} + c_2 { \fr{\l^2}{J^3}} + ... = J ( 1 +
c_1 \tl + c_2 \tl^2 + ...)$, where
 $J$ is the total $SO(6)$ spin $J= \sum_{i=1}^3 J_i$  and   $ \tl \equiv {\l \ov J^2}$
 is an effective semiclassical expansion parameter.
  $c_n= c_n({J_i \ov J})$ are functions of
  ratios of the spins which are finite in the semiclassical string-theory limit
$ J_i \gg 1$, $\tl =$fixed. Generic
 3-spin solutions are described by an
integrable Neumann model \ci{afrt,art}  and  the coefficients
$c_n$  are expressed in terms of genus-two
hyperelliptic functions.

Formally, string $\a' $  corrections
are suppressed in the limit $J \to \infty, \ \tl=$fixed
since $\a' = { R^2  \ov \sql}
\sim { 1 \ov J \sqrt{ \tl}}$. However, to   expect \ci{ft2}
to be able to compare these  classical energies  to the SYM
anomalous
dimensions \ci{mz1,bmsz,Min2}  one should
 check  that
 the $1 \ov J $ corrections are again
analytic in $\tl$  (as they are in the BMN case
\ci{ft1,par,cal}),   i.e. the expansion in
 large $J$ and small $\tl$
is well-defined on the string side,
\be \la{eee}
E= J \left[ 1 + \tl ( c_1 + { d_1 \ov J} + ...)
+  \tl^2 ( c_2 + { d_2 \ov J} + ...)  + ... \right] \ ,
\ \ \ \ \ \ \  \tl \equiv {\l \ov J^2}\ , \ee
with the classical energy  being the
$J\to \infty$ limit  of the exact expression.

This question was first addressed in \ci{ft3} on the example of
the simplest stable 3-spin solution of \ci{ft2}: a circular
string orbiting in $S^5$ with center of mass angular momentum
$J_1$  and two equal $SO(6)$  angular  momenta $J_2=J_3$ in the
two other  orthogonal planes. In addition to the values of
$J_1,J_2$ this solution is parameterized by an integer $\rk$
(winding
number) .\foot{Here we change the notation compared to
\ci{ft3}: there we had $J_1=J, \ J_2=J_3=J'$. Below $J$ will
stand for the total angular momentum  $J=J_1+ 2 J_2 $. Written in
terms of the $AdS_5$ time coordinate   $t$ and the angles of
$S^5$  (with the metric $ (ds^2)_{S^5} = d\g^2 + \cos^2\g\ d\vp_1^2
+\sin^2\g\ (d\psi^2 + \cos^2\psi\ d\vp_2^2+ \sin^2\psi\ d\vp_3^2)$)
the solution is \ci{ft2,ft3} (see also Appendix A): \  $ t= \k \tau ,
\ \g=\g_0 , \  \vp_1= \nu \tau ,
 \  \vp_2 =  \vp_3 =\ww \tau  ,  \  \psi= \kk \s $
 where $  \k,\g_0,  \nu, \ww$ are constants,  $\kk$ is an
 integer and  $
\ww^2 = \nu^2 +\kk^2 ,\
 \nu^2 = \k^2 -  2\kk^2 q,$ \ $q\equiv \sin^2\g_0$.
The  3   independent
 parameters  are  $\k$,  $q$  and  $\kk$.
The  non-zero  $SO(6)$
spin components are
$
J_1 =\sql\ \nu \  (1-q)$,\  $
J_2=J_3=  \ha \sql \ \sqrt{\nu^2 + \kk^2}\ q $.
The  classical
 energy  $E=\sql \k $   can  then be represented  as a function of the spins
 $E=E(J_2,J, \kk;\lambda)$.
 }
 At the classical (but not, in general,  at the quantum)  level
the dependence on $\kk$ can be absorbed into the
string tension $\sql$. Expanding $E$ one finds the explicit
form of $c_1$ in \rf{eee}
\ci{ft2,ft3}
\be
E_0=J+\fr{\l \rk^2J_2 }{J^2}+\cdots =\  J  \
\big( 1 + \tl \rk^2 {  J_2 \ov J}  + ... \big)  \ .
 \label{ce}
 \ee
The main reason to consider this solution (which is a special
case of a more general class of similar ``rational'' circular
3-spin solutions found in \ci{art})  is its simple analytic form
implying that the corresponding quadratic fluctuation action has
{\it constant}  coefficients (as in the BMN case corresponding to  the limit
$J_2=J_3=0$). As a result, the fluctuation frequencies which
determine the 1-loop correction to the energy (conjugate to the $AdS_5$ time
$t=\k \tau$)
 \bea
 E_1=\fr{1}{2\k}\bigg(\sum_{n\in Z}\omega_n^B
 -\sum_{r\in Z+\fr12}\omega_r^F\bigg) \label{e1}
 \eea
 can be readily found. Still,  $\omega$'s  are given \ci{ft3}
 by the
 roots of  {\it 4-order  polynomials}
 (see Appendix A)
 and thus are rather involved  functions of
  $J_2,J$ and $\kk$, making it difficult to compute the infinite
  sums in \rf{e1}.
  Attempting to evaluate (\ref{e1}) analytically,
  in \ci{ft3} the sums were
converted into integrals, but  it turns out that this  direct
procedure fails due to a singularity of the functions involved.

In this paper  we shall first improve  the general argument in
\ci{ft3} about the form \rf{eee} of the expansion of $E_1$ at
large $J$ and small $\tl$ and then   use numerical methods to
evaluate the first subleading  coefficient $d_1$.

A motivation behind this work is  to compare
the 1-loop string correction to the corresponding $1 \ov J$
correction in the  anomalous dimensions of the corresponding SYM
operators $\tr( \P_1^{J_1} \P_2^{J_2}  \P_3^{J_3}  ) + ...\  $.
On the gauge theory side,  one first expands in $\l$  and
then expands in $1 \ov J$,  so that  the
   anomalous dimensions  should  have the structure
   \be \la{dee}
\D= J + \l ( { a_1 \ov J} + { b_1 \ov J^2} + ...)
      + \l^2  ( { a_2 \ov J^3} + { b_2 \ov J^4} + ...) + ...\ .
       \ee
The  form of this expansion in the 2-spin ($SU(2)$) sector
was indeed verified  to first few leading orders in
 \ci{bmsz} and \ci{zar}.
 Moreover, it was checked on specific
 examples   \ci{bmsz,bfst,as,SS}   and also in general
  \ci{kru,kmmz,krt}  that the expressions for
 $a_1$  and $a_2$ match the coefficients $c_1,c_2$
 in $E$ \rf{eee}. Similar conclusion ($a_1=c_1$)
 was reached in the $SU(3)$ sector \ci{Min2,char,lopez,ST}
 (fluctuations near the circular 3-spin  solution of \ci{ft2}
 also match  \ci{frey}).

However, it was observed in \ci{SS,Min3} that disagreements start
at $\l^3$ order, $a_3 \not=c_3$,   with a plausible (``order-of-limits'')
explanation
suggested in \ci{SS,bds,afs}.
For that  reason,  it would be
interesting to see if the $1\ov J$ subleading  coefficient $b_1$ in
\rf{dee} agrees with the 1-loop coefficient $d_1$ in \rf{eee}. So
far,  $b_1$ was computed \ci{zar} only for a specific 2-spin  Bethe ansatz
state  corresponding to an  unstable state  on the string theory side
for which $d_1$  formally has an  imaginary part. In that case, a priori   the
comparison  does not seem to be well-defined.
%  (but see section 3 below).
Apart from clarifying this issue, it remains to compute $b_1$ for
the 3-spin state with $J_1 \not=0$, extending the  Bethe  ansatz
analysis of  \ci{Min2} where $a_1$ was found. Once this is done,
one will be in position to compare to the results  for $d_1$ on
the string side presented below.

An attempt of  comparison of our numerical result for $d_1$ in \rf{eee}
 for the 2-spin (unstable)  case with the
gauge theory result of \ci{zar} for $b_1$  indicates
a disagreement (see section 3).
We suspect that the  disagreement may  remain  also
in  the stable 3-spin case.  This seems  also to  suggest  that a similar ``1-loop''
(order $\l$)
disagreement  may be present   for  the $1\ov J^2$ correction  to
scaling dimensions of BMN operators.
An explanation  of these disagreements  may  be  again  related to  the non-commutativity  \ci{SS,bds}
of the ``string-theory''  (large $J$,   then small   $\tl$)  and the  ``gauge-theory''  (small $\l$,
then large $J$) limits.

\setcounter{footnote}{0}

%%%%%%%%%%%%%%%%%%%%%%%%%%%%%%%%%%%%%%%%%%%%%%%%%%%%%%%%%%%%%%%%%%%%%
\section{Structure of one-loop correction}
%%%%%%%%%%%%%%%%%%%%%%%%%%%%%%%%%%%%%%%%%%%%%%%%%%%%%%%%%%%%%%%%%%%%%

 In \cite{ft3} it was attempted to find the
  one-loop correction
$E_1$ in \rf{e1} as an expansion in  ${1\ov \kappa^2}= \tl + ... $, i.e.
%%%
 \bea \label{ea} E_1=\fr{1}{\k^2}e_1(q,
\rk)+\fr{1}{\k^4}e_2(q, \rk)+ ...
= \ \tl\  d_1({  J_2\ov J}, \rk)   + \ \tl^2\  d_2({  J_2\ov J}, \rk) + ...\ ,
 \eea
%%%
and
the expression for the leading order
coefficient $e_1$  was presented.
We used that \ci{ft3}
%%%
\be
 \fr{1}{\k^2}=\tl -\tl^2 { 2 \rk^2 J_2\ov J} + ... \ , \ \ \ \ \ \
 q\equiv \sin^2 \g_0 =  { 2 J_2 \ov J} + ... \ , \ee
%%%
 implying
%%%
 \be
 d_1= e_1 \ , \ \ \ \ \ \ \  d_2 = e_2 - \rk^2 q e_1 \ .
 \label{d1d2}
 \ee
%%%
However, later analysis  revealed that the functions that appear
at higher orders have unexpected irregularities, so that   the
method of
 \cite{ft3} needs a  modification.
 Here we shall briefly discuss the nature of this
modification (which turns out to be rather
 involved,  prohibiting a direct analytic computation)
  and then  turn
to numerical methods to evaluate $E_1$.

 It was
noticed  in \cite{ft3} that the bosonic and fermionic frequencies
(see  Appendix A below)
admit the following large $\k$ expansion
(with  ${n\ov \k}$ and ${r\ov \k}$  kept fixed)
 \bea \label{omb}
\om_{n}^B &=& \k\, \a_{-1}^B({n\ov \k})+ {1\ov \k}\a_{1}^B({n\ov
\k})
+ {1\ov \k^3}\a_{3}^B({n\ov \k})+ \cdots \ ,\\
\label{omf} \om_{r}^F &=& \k\, \a_{-1}^F({r\ov \k}) +
 {1\ov \k}\a_{1}^F({r\ov \k}) +
 {1\ov \k^3}\a_{3}^F({r\ov \k})+
\cdots \ . \eea
One can  think of $\alpha_a(\frac{m}{\k})$
as the values of functions $\alpha_a(x)$ at points $x_m={m\ov
\k}$. It was assumed in \cite{ft3} that all $\alpha_a(x)$'s are regular.
 In that case one could  replace the bosonic and
fermionic series in \rf{e1}
by integrals, and then
$\alpha_{2a+1}(\frac{m}{\k})$ with $ a\ge 1$ would not contribute
 to
the order $1\ov \k^{2}$ in the large $\kappa$ expansion.
However, it turns out that $\alpha^B_a(x)$  with $a \geq 3$
in general have singularities
(see Appendix B).
A more careful
analysis of  $\a_{2a+1}^B$ shows that at
small values of $x$ they behave as $\frac{1}{x^{2a}}+
{\cal{O}}(\frac{1}{x^{2a-2}})$. For this reason the analysis
 of  the large $\k$-expansion in \cite{ft3}
needs to be modified. One
must first subtract from
$\alpha_a(x)$   the singular contributions  and  after that
 convert the series into integrals. The
singular contributions cannot be represented by integrals and
have to be computed directly.
 However, then  all the terms of
the order $\frac{1}{x^{2a}}$ in $\a_{2a+1}^B$ with $a\geq 1$
contribute to $e_1(q, \rk)$ in \rf{ea}.
 For this reason,    obtaining the
complete answer for the coefficient $e_1(q,\rk)$ (and, in
general, for higher order coefficients
$e_p$) along these lines  would be hard
in practice.

In the $\kappa \rightarrow \infty$ limit the one-loop energy
correction must go to zero because this  strict limit is
essentially like a BPS limit - the bosonic and fermionic
contributions should then cancel against each other due to
supersymmetry.\foot{A  heuristic reason  is that in  the  strict
limit $\k \to \infty$ the world surface of the string becomes a
collection of  BMN  geodesics \ci{mik} with contribution of
tension between different string bits  effectively suppressed
(see also \ci{ft1} for a related  argument). } This
implies that only negative powers of $\k$ can appear in the large
$\k$-expansion of $E_1$. Indeed, examining the  functions
$\a_{a}(x)$ with $a > 1$  one can show that the one-loop
correction does have the large $\k$ expansion as given in \rf{ea}.

 To prepare the ground for  a  numerical
 evaluation of $e_1$ and $e_2$ in \rf{ea}
 let us first  discuss the convergence of the 1-loop
 correction \rf{e1} (which is expected  due to the conformal invariance
 of the underlying \adss string sigma model \ci{met}
 and can be demonstrated for a
 generic string solution,  see \ci{ft3}).
Each of the two   sums --    over the bosonic and the
fermionic frequencies --  is  divergent, and so
 one should
regularize them first.
Let us  use the standard  ``supersymmetry preserving''
regularization ($\epsilon \to 0$)
%%%
 \bea \label{E1reg}
 E_1=\fr{1}{2\k}\bigg(\sum_{n\in Z}e^{-\epsilon |n |}\ \omega_n^B
 -\sum_{r\in Z+\fr12}e^{-\epsilon |r |}\ \omega_r^F\bigg)\  .
\eea
%%%
Here $w_n^B$ ($w_r^F$) is the sum of  eight bosonic
(fermionic) frequencies at each level and
 as in \rf{e1}
$\omega$'s stand for their moduli, i.e. $+\sqrt{ \omega^2}$.
 One can then rearrange these sums as
 \bea \label{E1regi}
 E_1=\fr{1}{2\k}\sum_{n\in Z}\left(e^{-\epsilon |n |}\ \omega_n^B
 -\fr12 e^{-\epsilon |n- 1/2 |}\ \omega_{n-1/2}^F
-\fr12 e^{-\epsilon |n+1/2 |}\ \omega_{n+1/2}^F  \right)\ .
 \eea
%%%
This  can be further rewritten in a form which is more suitable
for taking the $\epsilon\to 0$ limit
 \bea
 E_1&=&\fr{1}{2\k}\bigg[ \sum_{n\in Z} e^{-\epsilon |n |}\Big(\omega_n^B
 -\fr12 \omega_{n-1/2}^F -\fr12 \omega_{n+1/2}^F  \Big) \nn \\
 &+& \fr12 \sum_{n\in Z}\left(
 e^{-\epsilon |n |}-e^{-\epsilon
|n- 1/2 |}\right)\omega_{n-1/2}^F +
\fr12\sum_{n\in Z}  \left(e^{-\epsilon |n
|}-e^{-\epsilon |n+1/2 |}\right)\omega_{n+1/2}^F
\bigg] \ . \label{E1regii} \eea
 %%%%
A  nice feature of (\ref{E1regii}) is that the series on the
first line is convergent even for $\epsilon =0$ because at large
$|n|$ (see Appendices A and B of \cite{ft3})
%%%
 \be \la{n3}
 \omega_n^B -\fr12 \omega_{n-1/2}^F
-\fr12 \omega_{n+1/2}^F\sim \frac{1}{|n|^3} \ .
 \ee
%%%
On the other hand,  the series in the second line of  \rf{E1regii}
can be easily computed in the limit $\epsilon\to 0$. First, we
note  that the fermionic frequencies $\omega_{r}^F$ are even
under $r\rightarrow -r$, and rewrite the second line of
\rf{E1regii}
 as
%%%
 \bea \label{E1regiii}
 \fr{1}{2\k}\left( e^{-\epsilon/2}+e^{+\epsilon/2}-2\right)
\sum_{r>0}e^{-\epsilon r }\omega_{r}^F\ .
 \eea
 %%%
Using the large $r$ expansion in \cite{ft3}
%%%
$$
\omega_{r}^F = 8 r + 4\;\frac{\k^2 - \rk^2 q}{r} +{\cal
O}(\frac{1}{r^3})\ ,
$$
%%%
%where $q=\sin^2 \g_0$ as defined in (\ref{q}),
we find that only
the first term, $8 r$, contributes in the limit $\epsilon\to 0$:
%%%
 \bea \label{E1regiv} \lim_{\epsilon\to 0}\left(
e^{-\epsilon/2}+e^{+\epsilon/2}-2\right) \sum_{r>0}e^{-\epsilon r
}\omega_{r}^F = 2\ .
 \eea
%%%
Thus, the  one-loop sigma model correction to the
 classical energy can be represented
 by the following convergent sum\foot{Let us stress
 again that one cannot formally rearrange the sum
without loosing the convergence.}  of the combination
 $\omega_n^B -\fr12 \omega_{n-1/2}^F -\fr12 \omega_{n+1/2}^F$
%%%
 \bea \label{E1regv}
 E_1=\fr{1}{2\k}\bigg[2+\sum_{n\in Z}\left(
 \omega_n^B -\fr12 \omega_{n-1/2}^F -\fr12 \omega_{n+1/2}^F \right)
 \bigg]\ .
 \eea
%%%
It is useful also to
 single out the
contribution of the $n=0$ term. Then
%%%
 \bea \label{E1K}
 E_1=\fr{1}{\k}\bigg[1+\fr{1}{2}\left(\omega_0^B
 - \omega_{1/2}^F \right)+\sum_{n=1}^N\left(\omega_n^B
 -\fr12 \omega_{n-1/2}^F -\fr12 \omega_{n+1/2}^F \right)
 \bigg] \ ,
 \eea
%%%
where we  have used the symmetry of the summand under $n\to
-n$. Having in mind a  numerical
computation of  $E_1$
we have also introduced the upper limit $N\to \infty$.
 The results discussed below were obtained for
$N=40000$.

%%%%%%%%%%%%%%%%%%%%%%%%%%%%%%%%%%%%%%%%%%%%%%%%%%%%%5
\section{Results of numerical evaluation}
%%%%%%%%%%%%%%%%%%%%%%%%%%%%%%%%%%%%%%%%%%%%%%%%%%%%%%%%

 Fixing the values of the parameters
 $(\k,q,\rk)$ one can, using Mathematica { or Maple},
 numerically solve the characteristic equations
 (\ref{bc}) and (\ref{fc})  for $\omega^B_n$ and $\omega^F_r$. The
 solutions are then substituted into (\ref{E1K}) to yield
 numerical values of
  \be
E_1=E_1(\k,q,\rk)\ .
 \ee
Since, in general, $E_1$  has a large  $\k$-expansion as given in
\rf{ea}, it is more convenient to compute not $E_1$ but $\k^2
E_1$. For large enough $\k$, the value of  $\k^2 E_1$ is very close to $d_1=e_1$, assuming
$e_2/\k^2$ is much smaller than $d_1$.

%Here  we shall present the numerical results of
%  the dependence
% of  $e_1,e_2$ on  $(q,\rk)$.
We consider  $\k=50,100,200$  for fixed values of $\rk=1,2,4,8$
and $q=\fr{h}{12}, \ h=0,1,2,...,12$  and for $\rk=1$ we set
$q=\fr{h}{24}$. Evaluating $\k^2 E_1$ numerically  it is possible
then to estimate $d_2$ to be sure it is small enough, and plot the
functions $d_1$. For example, we find:\
%%%
 $
 50^2 E_1(50,{1\ov 12},2)=0.2626  , \ \
 100^2 E_1(100,{1\ov 12} ,2)=0.2627   ,  \ \
 200^2 E_1(200, {1\ov 12},2)=0.2648  .
 $
%%%
Using  Mathematica Fit function then
yields the following $\k$-dependence,
%%%
 \be\la{fitt}
 \k^2 E_1 (\k, {1\ov 12} ,2)=0.2641-
    4.1752\fr{1}{\k^2} \ .
 \ee
%%%
We see that $d_2$ is  of order 1, and, therefore, $d_2/\k^2$ is
much smaller than  the value of $d_1=e_1$ which for this case is
$d_1 \approx 0.26$. The values of $N$ and $\k$ we used for the
computation do not allow us to find $d_2$ reliably because we
neglect the ``tail'' contribution ($\sum_{n=N+1}^\infty$) in
(\ref{E1regv}). It is shown in Appendix B that the tail
contribution to $\k^2 E_1$ is of order $\k^3/N^2$.  Because  of
that one cannot make $\k$ too large. For $N=40000$ and $\k = 200$
one has  $\k^3/N^2=0.005$, and, therefore, our computation is
accurate to at least  0.01.

This procedure can be repeated for the other values of $q\approx  { 2 J_2 \ov J} $ and
$\rk$. The resulting data is  shown in Table 1 and is  used to
plot the $q$-dependence of $d_1$ in Fig. 1-8.

%%%
\begin{table}[t]
\begin{center}
\begin{tabular}{|c|c|c|c|c|c|c|c|c|c|}
\hline
                & $\rk=1$       & $\rk=2$ & $\rk=4$ & $\rk=8$  \\
\hline
    $q=0$       &   -0.50       &  -1.00         & -2.00        & -4.00    \\
\hline
 $q=\fr{1}{12}$ &   -0.34       &  0.26          & 7.9       & 74  \\
\hline
 $q=\fr{2}{12}$ &   -0.20       &  1.33          & 16.1        & 133+ 5 i  \\
\hline
 $q=\fr{3}{12}$ &   -0.07       &  2.17          & 21.1+ 1.0 i  & 172+ 13 i  \\
\hline
 $q=\fr{4}{12}$ &    0.03       &  2.77          & 24.0+ 2.7 i  & 201+ 22 i  \\
\hline
 $q=\fr{5}{12}$ &    0.12       &  2.96          & 24.1+ 3.8 i  & 206+ 41 i \\
\hline
 $q=\fr{6}{12}$ &    0.17       &  2.47+ 0.77 i  & 22.3+ 8.0 i  & 192+ 67 i  \\
\hline
 $q=\fr{7}{12}$ &    0.18       &  1.93+ 1.27 i  & 18.6+ 11.0 i & 152+ 101 i  \\
\hline
 $q=\fr{8}{12}$ &    0.13       &  0.86+ 1.73 i  & 8.7+ 18.1 i  & 78+ 148 i  \\
\hline
 $q=\fr{9}{12}$ &   -0.23       & -1.77+ 3.21 i  & -7.7+ 31.0 i & -42+ 275 i  \\
\hline
 $q=\fr{10}{12}$ &  -0.28+ 0.50 i  & -1.49+ 6.47 i  & -8.6+ 57.5 i & -52+ 470 i \\
\hline
 $q=\fr{11}{12}$ &  -0.35+ 0.71 i &  -1.31+ 8.06 i  & -5.1+ 70.0 i & -24+ 575 i  \\
\hline
 $q=\fr{12}{12}$ &  -0.45+ 0.87 i  & -1.22+ 9.37 i  & -3.4+ 81.7 i & -9+ 673 i  \\
\hline
\end{tabular}
\end{center}
\caption{Values of $d_1(q,\rk)$}
\end{table}
%%%

It is important to  note that the circular string  solution in question is
stable, i.e. the  frequencies $\omega_n^B$  and thus $E_1$ are
real, only in the  following range of values  of $q$ (for fixed $\rk$):
 $q \leq  q_*$, where
$q_* = 1 - (1 - { 1 \ov 2 \rk})^2$  (see \ci{ft2,ft3} and Appendix
A).\foot{From numerical data in Table 1 we find, in agreement with
the analytic expression $q_* = 1 - (1 - { 1 \ov 2 \rk})^2$, that
$E_1$ becomes complex for $q > q_*$ where
 $
 \fr{18}{24}<q_*<\fr{19}{24} \;\;\;\mbox{for} \;\;\;\rk=1 $,
 $\fr{5}{12}<q_*<\fr{6}{12} \;\;\;\mbox{for} \;\;\;\rk=2$,
 $\fr{2}{12}<q_*<\fr{3}{12} \;\;\;\mbox{for} \;\;\;\rk=4$, and
 $\fr{1}{12}<q_*<\fr{2}{12} \;\;\;\mbox{for} \;\;\;\rk=8$.}
While  the plots are  valid  only   in the ``stable''
 regions of $q$,  we  have interpolated them to all values of $q \leq 1$
  by simply dropping the imaginary parts.

Let us first discuss the $q$-dependence of $d_1$ for $\rk=1$. The
plot of $d_1$ is shown in Fig.1.
\begin{figure}[!t]\centerline{
     \begin{minipage}[b]{10cm}    \epsfxsize=10cm
         \epsfbox{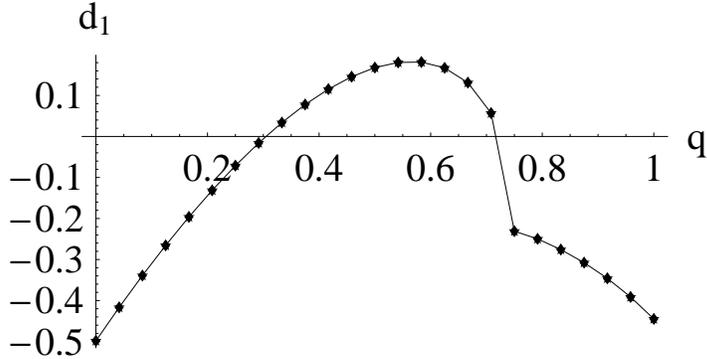}
      \end{minipage}
} \caption{$q$-dependence of $d_1$ for $\rk=1$} \label{e1k1}
\end{figure}
As discussed above the solution is stable for $q\le 0.75$. One can
see from the plot that the curve has a corner at $q=0.75$.  This
is a general property for all values of $\rk$ -- the energy is not
a differentiable function of $q$ at the edge of the stability
region. An interesting feature of the graph is that it crosses the
$q$-axis twice, at $q\approx 0.31$ and $q\approx 0.72$, i.e.   for
these values of $q$ the coefficient $d_1$ vanishes. \foot{Even
though the coefficient $d_2$ does not vanish at these points,
there exists  a curve $F(\lambda , q)=0$  on which  the first
(one-loop) sigma model correction, $E_1$,  vanishes. Thus, for the
corresponding
 values of $\lambda$ and $q$ the energy of the classical
circular string coincides with the exact energy of the quantum
string. This may be considered as a kind of non-renormalization
theorem  valid only for special  $\lambda$ and $q$
(for which we do not have any obvious explanation).}

\begin{figure}[!ht]
\centerline{
        \begin{minipage}[b]{10cm}
                \epsfxsize=10cm
                \epsfbox{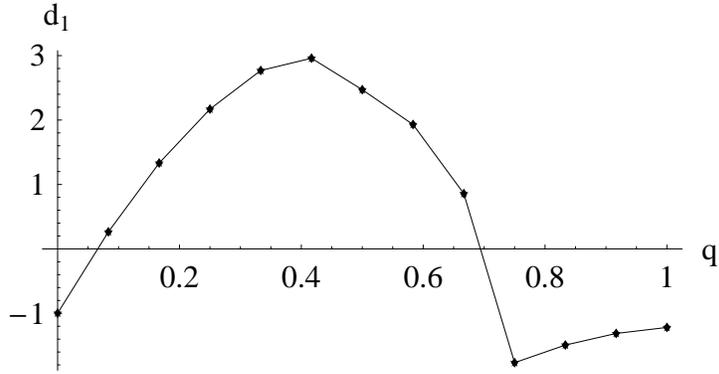}
        \end{minipage}
} \caption{$q$-dependence of $d_1$ for $\rk=2$} \label{e1k2}
\end{figure}
The plot of $d_1$ for $\rk=2$ is shown in Fig.2. In this case the
solution is stable for $q\le 0.4375$. One can see that the curve
crosses the $q$-axis only once in the stability region. To see
that the energy is not differentiable in this case either, it is
useful to plot $d_2$, see Fig. 3.
\begin{figure}[!ht]
\centerline{
        \begin{minipage}[b]{10cm}
                \epsfxsize=10cm
                \epsfbox{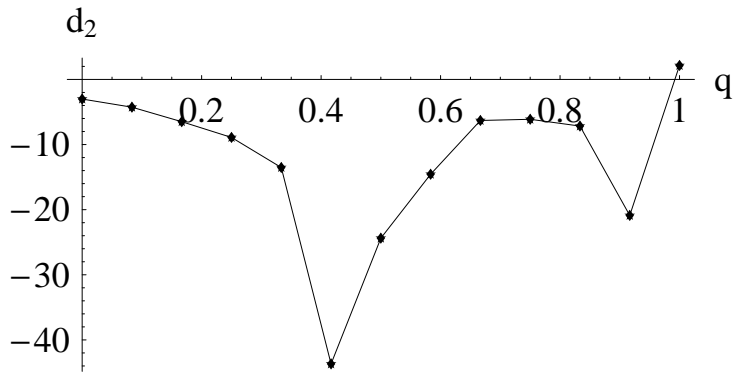}
        \end{minipage}
} \caption{$q$-dependence of $d_2$ for $\rk=2$} \label{eb2k2}
\end{figure}
Even though, as discussed above, the values of $d_2$ are not
reliable, one can clearly see that $d_2$ is not smooth at the edge
of the stability region.

The plots of $d_1$ for $\rk=4$ and $\rk=8$ are shown in Fig.4 and 5. The solution
is stable for $q\le 0.2344$ and $q \le 0.1211$, respectively.
\begin{figure}[!ht]
\centerline{
        \begin{minipage}[b]{10cm}
                \epsfxsize=10cm
                \epsfbox{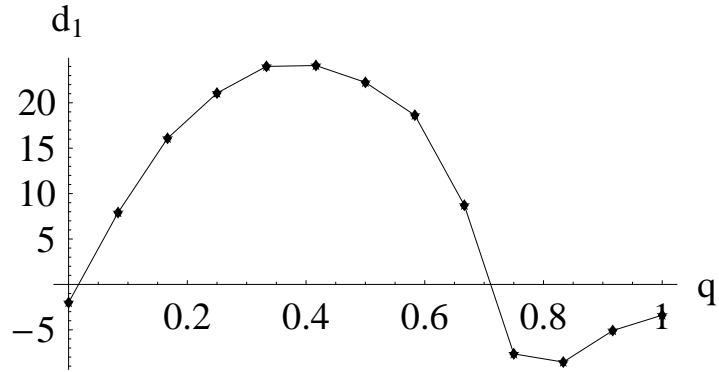}
        \end{minipage}
} \caption{$q$-dependence of $d_1$ for $\rk=4$} \label{e1k4}
\end{figure}
\begin{figure}[!ht]
\centerline{
        \begin{minipage}[b]{10cm}
                \epsfxsize=10cm
                \epsfbox{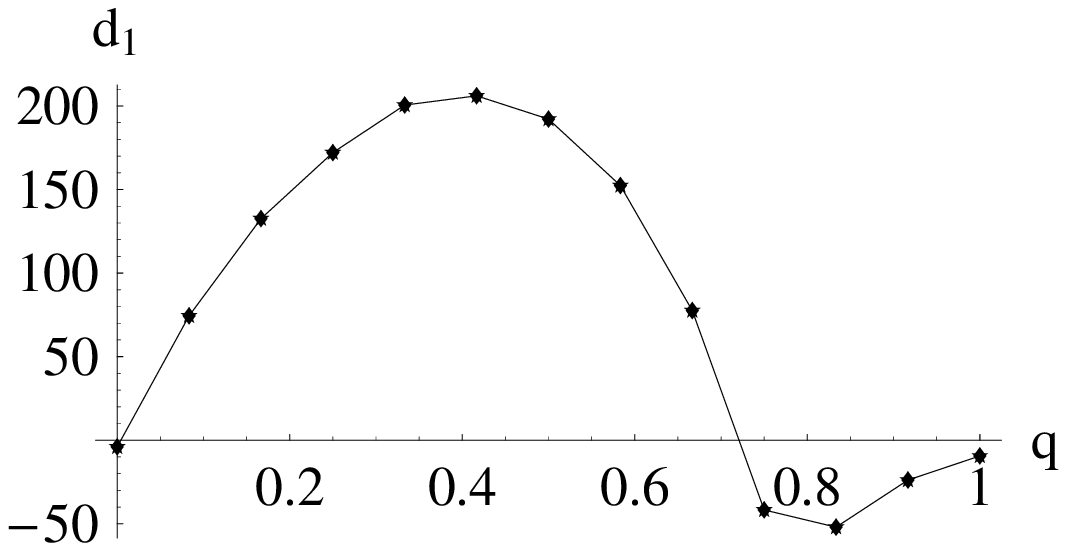}
        \end{minipage}
} \caption{$q$-dependence of $d_1$ for $\rk=8$} \label{e1k8}
\end{figure}
The plots of $d_1$ for all values of $\rk$ have similar shapes. In
particular, $d_1$ vanishes for at least   one value  $q=q_{\star}$
from the stability region. This  value  $q_{\star}$ depends on $\rk$,
suggesting that  one  should not expect to find a simple dependence of
$d_1$ on $\rk$ for fixed $q$.

Indeed, while   the leading  correction  in the classical energy  $E_0$ \rf{ce}
scales with $\rk$  as $\rk^2$, there is no  a priori reason why the leading coefficient in
$E_1$  should also have  a simple dependence on $\rk$.
The dependence of frequencies
$\omega^B_n$ and $\omega^F_r$
  on $\rk$  is such that    it can be eliminated by rescaling
  $\omega$'s, $\k$  and $n,r$ by $\rk$ \ci{ft3}
   (see also Appendix A), but
  since this transformation
   involves  a rescaling of  summation indices,
  the resulting $E_1$ \rf{e1} should, in general, be  a non-trivial function of $\rk$.
As follows from  Table 1, while  in the formal case of $q=0$
the dependence of $d_1$ on $\rk$ is  linear ($d_1(0,\rk) = -\ha \rk$),
it becomes non-trivial already for $q={1\ov 12}$ (see Fig. 6).
\begin{figure}[!ht]
\centerline{
        \begin{minipage}[b]{10cm}
                \epsfxsize=10cm
                \epsfbox{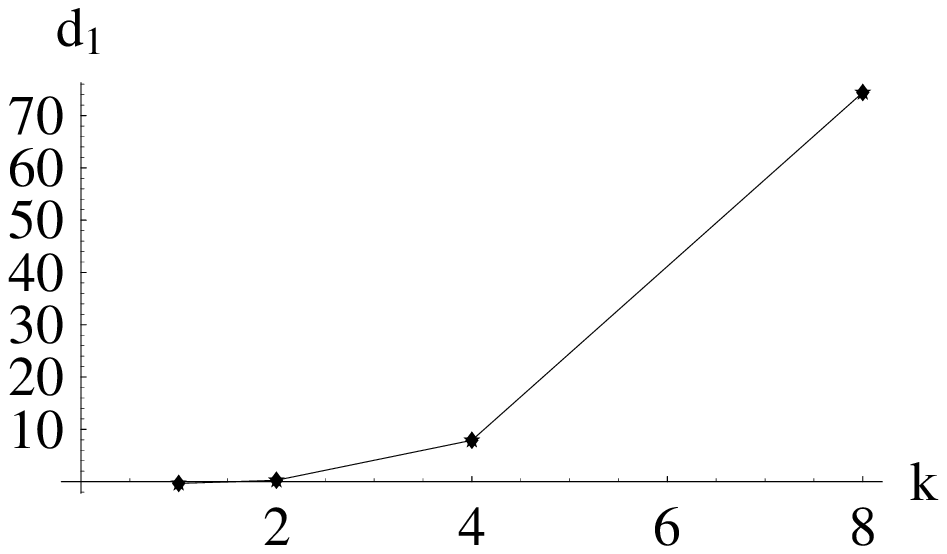}
        \end{minipage}
} \caption{$\rk$-dependence of $d_1$ for $q={1\ov 12}$}
 \label{kplot2}
\end{figure}
Since $d_1({ 1 \ov 12} ,1)=-0.34$ is negative and all other values
of $d_1({ 1 \ov 12},\rk)$ are positive, the $\rk$-dependence
cannot be given by a power function $\rk^p$ (but the curve can be
well-approximated by a polynomial of $\rk$). This example shows
that in general $d_1$ has a complicated dependence on   $\rk$. As
was mentioned  in the Introduction,  the one-loop gauge  theory
computation of the corresponding coefficient $b_1$ in \rf{dee}
was carried out   \ci{zar} only
  in  the (unstable)  $q=1$ case   with the result  $b_1 = \ha \rk^2$.
 {\it  If}   that gauge-theory prediction $ b_1 \sim  \rk^2 $  would  apply  also to the
  stable  $q < 1$ cases,
  then our results  would indicate a disagreement between the string-theory  ($d_1$)
and  the perturbative (1-loop) gauge theory  ($b_1$) coefficients.

Let us now consider
 the  special case of  $q=1$, i.e. $J_1=0, \   J_2=J_3= \ha J $
 in more detail.
 Here the
 frequencies can be found in a simple  analytic form
  \cite{ft3} and the computation of $E_1$
 becomes more explicit (see  Appendix B).
 The corresponding gauge-theory \rf{dee}  result
  \ci{zar}  written in the form \rf{eee}   reads
 \be \la{zaq}
 \Delta =\  J\  \bigg[ 1 + \ha \tl \rk^2  ( 1 + { 1 \ov J} + ...)
  + ...\bigg] \ . \ee
 Here the  leading-order term agrees
  \ci{bmsz}      with the classical string
 energy \rf{ce};     in order  for the ${ \tl  \ov J}$
 term in  \rf{zaq}  to be in agreement with
 the one-loop string correction in \rf{ea} one  should
 find that $d_1(q=1)  = \ha \rk^2 $.
 As already mentioned,   an  apparent  problem with checking this   is that
  the $q=1$  solution is   unstable for any   $\rk \geq 1 $ \ci{ft2,ft3}.
 In the simplest case of
  $\rk=1$
 there is one imaginary bosonic frequency (for larger  $\rk$
 there are several unstable modes).
 As a result,  the definition and interpretation
 of the 1-loop correction to the energy becomes  non-trivial
 (formally, the 1-loop correction then  contains  an  imaginary part
 determining the rate of decay
 of the unstable state, see, e.g., \ci{wei}).
 In order  to see if  string theory result may be put
 into an  agreement with the gauge theory result \rf{zaq}
 we  may try to use one  of the following definitions of $E_1$
 (for definiteness, we shall  consider the case  of $\rk=1$):\foot{
Formally,    the unstable   mode  is not ``seen'' on the gauge theory side \ci{bmsz}.
 More precisely, an unstable mode of the  3-spin circular solution
 corresponds  to the configuration  when  a  Bethe root  moves  off the  real axis
 \ci{frey,Min3};
  this case does not correspond to a true eigenstate of the
 hermitian  Hamiltonian of the Heisenberg ferromagnet \ci{mz1}.
 Still, this suggests that some analytic continuation  may  apply.
 Related aspect of this  problem is
 that the spin chain states found using
 the Bethe ansatz \ci{bmsz} are exact quantum states, while
 on the string theory side we are considering semiclassical states dual
 to coherent states of the spin chain \ci{kru,krt,kt}.
 One can show (see also \ci{lopez})
  that the corresponding unstable mode is present also
  in the
 ``Landau-Lifshits''   sigma model  which is
  the coherent state effective
 action following in the low-energy approximation
 from the Heisenberg spin chain Hamiltonian and which agrees
\ci{kru} with a
 large spin limit of the string sigma model action.}

 (i) compute $E_1$ as a sum over all frequencies as in  \rf{e1},\rf{E1K}
 and omit the imaginary part;  this  amounts to ignoring  the
 contribution  of the one unstable bosonic mode with $n=1$.

 (ii)  analytically continue the  value of the mass of the
 ``tachyonic''  $n=1$ mode,
 i.e. include  the contribution  of  its   frequency
 to $E_1$ with the modulus sign.\foot{
A  possible way to support the second prescription
 is to view the $J_2=J_3, \ J_1=0$ case  as an analytic continuation of the
stable solution  with $J_2=J_3, \  J_1 >  { 2 \ov 3}  J_2$. For this stable
solution the  spectrum of fluctuations (and thus $E_1$)  is
 real
and matches with Bethe ansatz $J=\infty$ spectrum \ci{Min2}. Then
we may  analytically continue  all relations in the
angular momentum  plane   and try to  define the  $J_1 \to 0$
limit. }

 In the  case (i)  we find by the  numerical evaluation of the sum
 that
 $ d_1 \approx -0.446  $.
 In the case (ii) we get instead $ d_1 \approx 0.42$ (the additional
 contribution of the modulus of the frequency of the  unstable $n=1$
 mode is $\delta d_1 = { \sqrt 3 \ov 2} \approx  0.866$).
 This  may look    close to the $0.5$ value in \rf{zaq},   but our estimate of the
 numerical error is much smaller than $0.08$, so we
 are inclined  to conclude that there is a disagreement between the gauge theory
 and string-theory values for $d_1$,  with  a  plausible explanation
 being the same order-of-limits problem  as in \ci{bds}
 (see also  section 4 below).

Finally, let us comment on the  $\rk$-dependence of $d_1$   in this  $q=1$ case.
Following the first prescription (i), i.e.   keeping only
the real part of $d_1(\rk) $ we get the plot
 in Fig.\ref{kplot}. It is interesting to note
that the curve can be well-approximated by the power function
$-0.446\, \rk^p$ with $p= 1.46$.
\begin{figure}[!ht]
\centerline{
        \begin{minipage}[b]{10cm}
                \epsfxsize=10cm
                \epsfbox{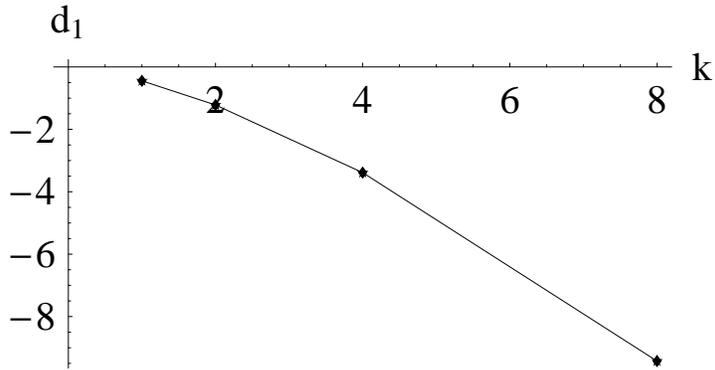}
        \end{minipage}
} \caption{$\rk$-dependence of the real part of $d_1$ for $q=1$}
\label{kplot}
\end{figure}

With  the second prescription (ii), i.e.  taking  the sum of the
real part and  the imaginary part\foot{All  unstable modes happen to
have purely-imaginary frequencies, so their modulus is equal to
the absolute value of their imaginary part.}
 of $d_1$  we get the plot in  Fig.\ref{kplotii}.
\begin{figure}[!ht]
\centerline{
        \begin{minipage}[b]{10cm}
                \epsfxsize=10cm
                \epsfbox{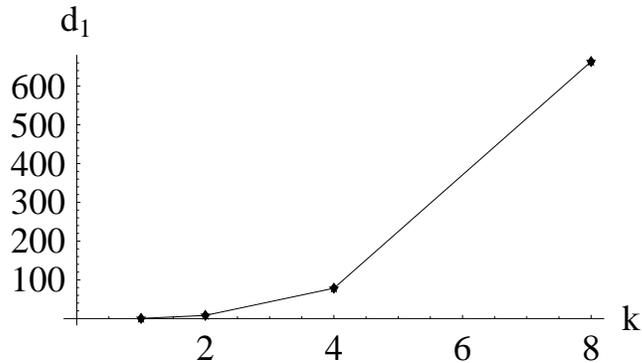}
        \end{minipage}
} \caption{$\rk$-dependence of  $d_1$ for $q=1$ in the absolute value prescription}
\label{kplotii}
\end{figure}
In this case it cannot be approximated by a power function.
Comparing to \rf{zaq} suggests again that  there is a
 disagreement
between the $1/J$ string theory and gauge theory results.

%%%%%%%%%%%%%%%%%%%%%%%%%%%%%%%%%%%%%%%%%%%%%%%%%%%%%5
\section{Concluding remarks}
%%%%%%%%%%%%%%%%%%%%%%%%%%%%%%%%%%%%%%%%%%%%%%%%%%%%%%%

In this paper we have used numerical methods to
 analyze
the leading 1-loop sigma model correction to the energy of the
classical circular spinning string \ci{ft2}.  We have confirmed
the expected large $J$ expansion of the energy, and studied the
dependence of the first sub-leading coefficient $d_1$ in \rf{eee}
on the two parameters -- $q\approx { 2 J_2 \ov J} $ and  ``winding
number'' $\rk$. Comparing our results with the known gauge-theory
result  \ci{zar} for the corresponding spin chain state (dual to
unstable 2-spin circular string with  $q=1$), we have found a
discrepancy not only in the numerical value but also in the
$\rk$-dependence of the leading 1-loop coefficient $d_1$. Even
though our computation is unambiguous and reliable only for the
stable 3-spin string states  with $q  \leq   q_* < 1$, the
different $\rk$-dependence  of $d_1$  may be viewed as an
indication  that there is a string/gauge theory disagreement  for
the $1/J$ correction starting already at one-loop order on the
gauge  theory side. We should add a reservation  that
 it is still possible that the disagreement we find
  is due to the fact that  the semiclassical quantization
   and its comparison to the gauge theory side
  are not  directly
  applicable  in the case of  the unstable 2-spin solution, and
  there is still a chance that one
   may find a one-loop agreement for the stable 3-spin string
   states  once one computes the corresponding
   $1/J$ gauge theory  corrections  using  the $SU(3)$  Bethe ansatz  of \ci{Min2}.

Assuming this $1/J$ disagreement persists, it   should  have the same origin as
 the  previously found
mismatch between the  string and gauge theory results at  3-loop
order in $\l$ and the leading order in $J$ \ci{SS}. As was
suggested in \ci{SS,bds}, the  latter   can be explained  by
adding ``wrapping''   contributions to the dilatation operator
(and thus to the Bethe ansatz relations) on the gauge theory side.
For example, in  the $q=1$ case  one may use the function like $
\l^J/(1 + \l)^J$ which is 1 in the string theory  limit ($J\to
\infty$ with fixed $\l/J^2 \equiv \tilde \lambda$)  but zero in
the perturbative gauge theory limit to interpolate between the
different ${\l^3  \ov J^5}$  results as follows:
$$ \Delta = J + { \l \ov 2 J}  - { \l^2 \ov 8 J^3} +
   { \l^3  \ov 16  J^5} { \l^{J-3} \ov (1 + \l)^{J-3}}  + ... \ .
$$
This expression  agrees with both the string ($E=\sqrt{J^2 + \l}$ \ci{ft2})  and the
perturbative gauge theory ($\Delta_{\rm pert}=   J + { \l \ov 2 J}  - { \l^2 \ov 8
J^3} + 0\times  \l^3 + ...$  \ci{SS})  results.
Same idea may be applied to explain the discrepancy at order $\l \ov J^2$:
for example, if we assume that the interpolation formula contains also the term
$$  \Delta = ... +  { \l  \ov 2 J^2}  \big[1 +    a     { \l^{J-1} \ov (1 +
\l)^{J-1}} \big]   + ... \  ,
$$
then the   gauge theory limit   result for the coefficient $d_1$ will be $\ha $
as in \ci{zar}  while the string theory limit  will give
$d_1 =\ha (  1 + a)  $, explaining  the apparent disagreement of our
result with that of \ci{zar}.

A related observation is that this apparent $1/J$   disagreement
can be easily accommodated  and thus explained   within   the
generalized Bethe ansatz for quantum string spectrum  recently
proposed in \ci{afs}. To this end all one should do is to assume
a definite large $L\equiv J$ expansion of the functions
$c_r(g,L)$\  ($g\equiv \sqrt{8 \pi^2 \lambda}$)  appearing  in
the Bethe ansatz of \ci{afs}. In particular, one can see that if
$c_r$ have an expansion of the form  $c_r(g,L) \sim \tl^{r+2}
L^{2r +4} + \tl^{r} L^{2r +3} $,  then at order $1/L\equiv 1/J$
there is a string/gauge theory mismatch already for the
coefficient of the
 one-loop ($\sim \tl$) term.
   The
Bethe ansatz  of \ci{afs} also implies that if there is a string/gauge theory
(dis)agreement for spinning string states  at order $1/J^n$ then a
similar (dis)agreement should exist also  for the BMN states at
 order $1/J^{n+1}$.  In view of the above discussion,
this suggests  that
%a near plane wave computation of the energy
for the  BMN states   the  disagreement at order $1/J^2$ should start not  for 3-loop
($\l^3$) terms as at  $1/J$  order but already for the 1-loop ($\l$)  terms.

  %this specific conjecture assumes that
 %on gauge theory side the wrapping term starts as
%$ \lambda^J (  1/J^3 +  1/J^5 + ... )$  this is rather strange
%(and second  term takes care of 3-loop classical disagreement while the first
% --  of our $1/J $  disagreement)
 %but may be there is a way of connecting this conjecture with $1/J^2$
% disagreement in near-BMN case... though I am not sure  yet how.
% in any case $c(1/J)= 1/J^k$  coeff of $\lambda^J$   term seems
%a plausible explanation of disagreements with naive gauge theory
%perturbative results
%it may be more complicated, all we are saying is that there are
%$J$-dependent coeffs  of terms  $\lambda^J + \lambda^{J+1} + ...$
%and then we resum perturb expansion in lambda.

\setcounter{section}{0}
%%%%%%%%%%%%%%%%%%%%%%%%%%%%%%%%%%%%%%%%%%%%
\setcounter{equation}{0}
\setcounter{footnote}{0}
%%%%%%%%%%%%%%%%%%%%%%%%%%%%%%%%%%%%%%%%%%%%
\vspace{0.2in}
%%%%%%%%%%%%%%%%%%%%%%%%%%%%%%%%%%%%
{\bf Acknowledgments}
%%%%%%%%%%%%%%%%%%%%%%%%%%%
\vspace{0.1in}

We are grateful to N. Beisert  and 
 K. Zarembo  for useful  discussions. Part of the work was done
 while S.F. visited SPHT/Saclay. He is grateful to I. Kostov and V. Schomerus for
 warm hospitality.
 The work of S.F. was supported in part by the 2004 Crouse Award.
 The work of I.P.  and
A.T. was supported  in part by the  DOE grant DE-FG02-91ER40690.
I.P is grateful to S. Mathur for his support and encouragement.
A.T.   acknowledges the hospitality of KITP at Santa Barbara
during the completion of this work where his research was partly supported by
the NSF under grant No. PHY99-07949.
A.T. is also supported  by the INTAS contract 03-51-6346 and the RS Wolfson award.

\newpage

\appendix{Classical solution  and
 quadratic \\  fluctuations}
%%%%%%%%%%%%%%%%%%%%%%%%%%%%%%%%%%%%%%%%%%%%%%%%%%%%%%
\renewcommand{\theequation}{A.\arabic{equation}}
\setcounter{equation}{0}

The solution we discussed above was found in \ci{ft2}, and the
characteristic equations for the  quadratic fluctuations near it were
obtained in \cite{ft3}. Here we briefly review the derivation of
the bosonic characteristic equation while in   the fermionic
case  we only quote the final result referring
to \cite{ft3} for more details.
The bosonic part of the string action in the conformal gauge is
 $I= {\sqrt{\l}}\int d\tau \int_0^{2\pi} \fr{d\s}{2\pi}
 \left(L_{AdS}+L_S\right)$, where
%%%
 \bea
 L_{S}&=&-\fr12 \pa_a X_{M} \pa^a X_{M}
 -\fr12 {\Lambda}( X_{M}X_{M}-1)\ ,  \label{action}\\
 L_{AdS}&=&-\fr12 \eta^{PQ}\pa_a Y_{P} \pa^a Y_{Q}
 +\fr12 \tilde{\Lambda}(\eta^{PQ} Y_{P}Y_{Q}+1)\ .
 \eea
%%%
Here $X^M,\  M=1,...,6$ and $Y^P,\ P=0,...,5$ are the embedding
coordinates with a flat Euclidean metric for $S^5$  and with
$\eta_{MN}=(-1,+1,+1,+1,+1,-1)$ for the  $AdS_5$ respectively.
We consider the  configuration where the string is located in
the center of $AdS_5$ while rotating in $S_5$. The $AdS_5$ part
of the solution is trivial ($ Y_5+iY_0=e^{it}, \ Y_1,...Y_4=0$),
%%%
with the global $AdS_5$ time  being set to $t=\k \tau$, while the
$S^5$ part is
%%%
 \bea
 X_1+iX_2=\sqrt{q}\  \cos k\s\; e^{i\ww\tau}, \
 X_3+iX_4=\sqrt{q} \ \sin k\s\; e^{i\ww\tau}, \
 X_5+iX_6= \sqrt{1-q} \;e^{i\n \tau} \label{jjsol}
 \eea
%%%
with
%%%
 \be
 \mbox{w}^2=\nu^2+\rk^2\ , \ \ \ \ \ \ \    \ \L=\nu^2 \ , \ \ \ \ \ \ \
\nu^2 = \k^2 -  2\kk^2 q \ ,\ \ \ \ \ \ \ q\equiv \sin^2 \g_0.
 \ee
%%%
It was shown in \cite{ft3} that the quadratic fluctuation
Lagrangian around this solution can be written as
%%%
 \bea
 L_2 =&&  (\del_\tau \bx_s )^2    - (\del_\s \bx_s)^2
 + 4 \nu \sqrt{q}\   \bx_5   \del_\tau \bx_6
- 4 \ww ( \sqrt{1-q}\  \bx_5  \del_\tau \bx_2  -  \bx_3\del_\tau
\bx_4) \nn\\
&& { + \  4 \kk(\sqrt{1-q}  \  \bx_5     \del_\s \bx_3   -  \bx_2
\del_\s \bx_4)  \ .  } \label{lfinal}
 \eea
%%%
The corresponding fluctuation  spectrum
is found by using the following mode
expansion
%%%
 \be
 \bx_s = \sum_{n=-\infty}^{\infty} \sum_{h=1}^8  A^{(h)}_{sn}\
e^{i( \om_{n,h} \tau  \ + \  n \s)} \ ,
 \ee
%%%
where $h$ labels the different frequencies for a fixed value of
$n$. The determinant of  the characteristic matrix
 is proportional to (here we set $\Om
\equiv \om^2_n$)
 \bea
  B_8(\Om)  =
  &&  {\Om}^4+{\Om}^3\,( -8\,\rk^2 - 4\,n^2
    + 20\,\rk^2\,q -8\,{\k}^2  )
     \nn\\
 &&  +
   {\Om}^2\, (16\, \rk^4+32\, \rk^2 \k^2+16\,\k^4+8\,\rk^2 n^2
        + 16\,\k^2 n^2+6\,n^4-80\,\rk^2\k^2 q\nn\\
  &&\hspace{1in}     -80\,\rk^4 q  -36\,\rk^2n^2q+96\,\rk^4 q^2  ) \nn\\
   && + \Om\, (-32\,\rk^4 n^2-32\,\rk^2 \k^2 n^2+8\,\rk^2 n^4-8\k^2 n^4
       -4\,n^6+96\,\rk^4 n^2 q  \nn\\
       &&\;\;\;\;\;\;\;\;\;\;\;+48\,\rk^2 \k^2 n^2 q+12\,\rk^2 n^4 q
       -96\,\rk^4 n^2 q^2) \nn \\
    && + 16\,\rk^4\,n^4 - 8\,\rk^2\,n^6 + n^8  -
  16\,\rk^4\,n^4\,q    +
  4\,\rk^2\,n^6\,q
 \label{bc}
 \eea
%%%
The 4+4  $S^5$-frequencies are obtained as $\omega_{n,S^5}^B=\pm
\sqrt{\Omega}$ where $\Omega$ is one of the  4 roots of $B_8=0$. In
addition,  there are 4+4
 $AdS_5$ frequencies
%%%
 $
 \om_{n}=\pm \sqrt{ n^2 + \k^2 }
. $
%%%
By following the analogous  steps
one can show  \cite{ft3}  that the fermionic
characteristic frequencies $\omega_r^F$  are determined by $F_8(\Om)=0$ where
 %%%
 \bea
 F_8=&& 2\,\Om^4 +
  {\Om}^3\,\left( -8\,\rk^2 - 12\,{\k}^2 - 8\,r^2 +
     20\,\rk^2\,q \right) \nn\\
 &&     +
  {\Om}^2\,\left( 12\,\rk^4 + 28\,\rk^2\,{\k}^2 +
     18\,{\k}^4 + 8\,\rk^2\,r^2 + 28\,{\k}^2\,r^2 +
     12\,r^4 - 52\,\rk^4\,q \right. \nn\\
    &&\left.\;\;\;\;\;\;\;\;\;\;\; -
     64\,\rk^2\,{\k}^2\,q -
     36\,\rk^2\,r^2\,q + 59\,\rk^4\,q^2 \right)
     \nn\\
 &&    +
  \Om\,\left( -8\,\rk^6 - 20\,\rk^4\,{\k}^2 -
     20\,\rk^2\,{\k}^4 - 8\,{\k}^6 + 8\,\rk^4\,r^2 +
     8\,\rk^2\,{\k}^2\,r^2 - 20\,{\k}^4\,r^2 \right. \nn\\
     &&\left.\;\;\;\;\;\;\;\;+
     8\,\rk^2\,r^4 - 20\,{\k}^2\,r^4 - 8\,r^6 +
     44\,\rk^6\,q +
     80\,\rk^4\,{\k}^2\,q +
     44\,\rk^2\,{\k}^4\,q \right.\nn\\
     &&\left.\;\;\;\;\;\;\;\; -
     24\,\rk^4\,r^2\,q +
     32\,\rk^2\,{\k}^2\,r^2\,q +
     12\,\rk^2\,r^4\,q - 78\,\rk^6\,q^2 -
     79\,\rk^4\,{\k}^2\,q^2\right.\nn\\
     &&\left.\;\;\;\;\;\;\;\;\;+
     2\,\rk^4\,r^2\,q^2 + 45\,\rk^6\,q^3
     \right)\nn\\
    && +2\,\rk^8 + 4\,\rk^6\,{\k}^2 + 2\,\rk^4\,{\k}^4 -
  8\,\rk^6\,r^2 - 4\,\rk^4\,{\k}^2\,r^2 -
  4\,\rk^2\,{\k}^4\,r^2 + 12\,\rk^4\,r^4 \nn\\
  &&-
  4\,\rk^2\,{\k}^2\,r^4 + 2\,{\k}^4\,r^4 -
  8\,\rk^2\,r^6 + 4\,{\k}^2\,r^6 + 2\,r^8 - 12\,\rk^8\,q -
  16\,\rk^6\,{\k}^2\,q \nn\\
  &&-
  4\,\rk^4\,{\k}^4\,q +
  28\,\rk^6\,r^2\,q +
  16\,\rk^4\,{\k}^2\,r^2\,q +
  4\,\rk^2\,{\k}^4\,r^2\,q -
  20\,\rk^4\,r^4\,q\nn\\
  && + 4\,k^2\,r^6\,q +
  27\,\rk^8\,q^2 +
  21\,\rk^6\,{\k}^2\,q^2 +
  2\,\rk^4\,{\k}^4\,q^2 -
  30\,\rk^6\,r^2\,q^2\nn\\
  && -
  11\,\rk^4\,{\k}^2\,r^2\,q^2 +
  11\,\rk^4\,r^4\,q^2 - 27\,\rk^8\,q^3 -
  9\,\rk^6\,{\k}^2\,q^3 \nn\\
 && +
  9\,\rk^6\,r^2\,q^3 + \frac{81\,\rk^8\,q^4}{8}\ .
 \label{fc}
 \eea
 %%%
Unlike the bosonic case,  the $AdS_5$ and $S_5$
parts are not decoupled
 in the fermionic case. The eight  fermionic frequencies are obtained by
 solving $F_8=0$ and taking $\omega^F=\sqrt{\Om}$ with double
 degeneracy.

 When  solving  $B_8=0$  one may set $\rk=1$;
 then the $\rk$-dependence can be restored by the
 following rescaling,
 %%%
 \bea
 \omega_n\rightarrow \fr{\omega_n}{\rk},\;\;\;\ \  n\rightarrow \fr{n}{\rk},
 \;\;\;\ \ \k\rightarrow \fr{\k}{\rk}\  .
 \eea
 %%%
 Similar  rescaling  can be  done  in   the fermionic case \ci{ft3}.

Let us now consider the large $\k$-expansion
of the bosonic frequencies to analyze
the stability
condition in that limit
%%%
 \bea
 \omega_n^2\ \ \rightarrow\ \  \fr{h_0}{4\k^2}+\fr{h_1}{\k^4}+\cdots \ .
 \eea
%%%
Here
%%%
 \be
 h_0=n^2 \big[ 2\,\rk^2 (2 - 3 q) +n^2
 \pm 2\rk \sqrt{4\, n^2( 1 - q)     +     \rk^2 q   (9q - 8 )    } \big]
 \ee
%%%
The stability condition that follows from
positivity of $h_0$  is \ci{ft3}
%%%
 \be
 q\leq q_* \ , \ \ \ \ \ \ \
 \ \  \  q_* =  1- ( 1 - \fr{1}{2\rk})^2 \ .  \label{sca}
 \ee
%%%

%%%%%%%%%%%%%%%%%%%%%%%%%%%%%%%%%%%%%%%%%%%%%%%%%%%

\appendix{$q=1$ case:  \ \ $J_1=0, \ J_2=J_3$}

%%%%%%%%%%%%%%%%%%%%%%%%%%%%%%%%%%%%%%%%%%%%

%%%%%%%%%%%%%%%%%%%%%%%%%%%%%%%%%%%%%%%%%%%%%%%%%%%%%%
\renewcommand{\theequation}{B.\arabic{equation}}
\setcounter{equation}{0}

In the case of $q=1$ the  string  is stretched around the big circle of $S^5$
and rotates about  its   center of   mass with two equal angular momenta.
Here
 the characteristic equations
\rf{bc} and \rf{fc} can be solved explicitly
\cite{ft3} and one finds that the bosonic
 $S^5$ frequencies   are
 (up to an overall sign change)
%%%
 \bea
 \w_{n}^B &=& \big[n^2+2\k^2-2\rk^2
 \pm 2\sqrt{\left(\k^2-\rk^2\right)^2+n^2\k^2} \  \big]^{1/2 }
 \nn \\
 \w_n^B &=&  \sqrt{n^2+\k^2-2\rk^2}
  \pm\sqrt{\k^2-2\rk^2}
 \ .
 \label{wb}
 \eea
%%%
This may be compared to the  $AdS_5$ fluctuation
 frequencies
%%%
 $
 \w_n^B= \sqrt{ n^2+\k^2 }$.
%%%
 The fermionic frequencies are (with double degeneracy)
%%%
 \bea
 \w_r^F=  \fr12\left(2\sqrt{r^2+\k^2-\rk^2}\pm \sqrt{\k^2-\rk^2}
      \pm \sqrt{\k^2-2\rk^2} \right)\ .  \label{wf}
 \eea
 Using (\ref{wb}) and (\ref{wf})  we get  the
explicit form of (\ref{E1K}) is
%%%
 \be
%&& \fr{1}{\k}\sum_{n=1}^N\left(w_n^B
% -\fr12 w_{n-1/2}^F-\fr12 w_{n+1/2}^F\right)\nn\\
 E_1={1\ov \k} \bigg[1+  \big(\ \sqrt{\k^2-2\rk^2 }+\sqrt{\k^2-\rk^2 }
+  2 \k-  4 \sqrt{{\textstyle { 1 \ov 4}}+\k^2-\rk^2} \  \big)
 +\sum_{n=1}^N  \S (n,\k, \rk)\bigg]
    \ ,\la{kaE1} \ee
%%%
 where
 \bea \la{sss}
\S  =&& \sqrt{  ( n  + \sqrt{n^2 - 4 \rk^2   } )^2        + 4 \k^2  }
+{2}\sqrt{n^2-2\rk^2+\k^2  }
 +{4}\sqrt{ n^2+\k^2 } \nn\\
 &&
-{4}\sqrt{(n-1/2)^2-\rk^2 +\k^2}-{4}\sqrt{(n+1/2)^2-\rk^2 +\k^2}
\;\;
 \eea
%%%
We have used that
$$
\big[   n^2+2\k^2-2\rk^2
 + 2\sqrt{ (\k^2-\rk^2)^2 +n^2\k^2 } \big]^{1/2}   +
 \big[ n^2+2\k^2 -2\rk^2
 -2\sqrt{ (\k^2-\rk^2)^2
 +n^2\k^2} \big]^{1/2} $$ \be
 = \big[2 (n^2+2\k^2- 2 \rk^2)
 + 2 n \sqrt{n^2 - 4 \rk^2   } \big]^{1/2} \  . \ee
Note also that
$\S_{\k \to \infty}  \to { 1 \ov 2 \k} [ 2 k^2 - 2 - n^2    +
 n \sqrt{n^2 -4k^2 }] + O({1 \ov
\k^3})$ and
$\S_{n \to \infty}  \to {k^2 - 2 k^4  - \k^2  \ov n^3}  + O({1 \ov
n^5})$, in agreement  with \rf{n3}.

One may be  tempted to evaluate the  sum  in \rf{kaE1} by
converting it into an integral. This  conversion is not possible,
however, due to singularities of the  summand. To see this let us
set $\rk=1$ and follow the procedure of \cite{ft3} to  perform the
large $\k$-expansion of $\S$ for  fixed  $\fr{n}{\k}\equiv x$,
%%%
 \bea
\S(x) \equiv  \S ({\k x}, \k, 1) \ { \to _ {_{\k \to \infty} }}
 -\fr{1}{(1+x^2)^{3/2}}\;\fr{1}{\k}
 -\fr{\;\;16+71x^2+76x^4+16x^6}{16x^2(1+x^2)^{7/2}}
 \;\fr{1}{\k^3}+ ...   \ .
 \eea
%%%
Note that the first term in the coefficient of $\fr{1}{\k^3}$ has
a singularity at $x=0$. Even though we singled
 out the zero mode
contribution in (\ref{E1K}), it is this divergent behavior near
$x=0$ that is responsible for the failure of the conversion of the
sum  to an integral.

To estimate the accuracy of the numerical method that we used
in section 3 we
approximately evaluate the tail of the sum in (\ref{kaE1}). It is
 possible to convert the tail part of the sum in (\ref{kaE1}),
i.e., ${1 \ov \k} \sum^\infty_{{N+1}} {\S}$,
to an integral over $x = {n \ov \k}$ since it
does not contain $x=0$ or its neighborhood.
 Let us follow \cite{ft3} and use that (we will not distinguish
 between $N$ and $N+1$ since $N \gg 1$)
%%%
 \bea
\fr{1}{\k} \sum^\infty_{n=N}\;\S \approx
\int^\infty_{\fr{N}{\k}}dx\;
\ g(x)+O(\fr{1}{\k^6})\ ,
 \label{gvint}
 \eea
%%%
where
%%%
 \bea
 g(x)=-\fr{2}{15}\S (x-\fr{1}{\k})+\fr{6}{5}\S(x-\fr{1}{2\k})
 +\fr{1}{30}\S(x)-\fr{2}{15}\S(x+\fr{1}{2\k})+\fr{1}{30}
 \S(x+\fr{1}{\k})\ ,
 \eea
%%%
and  $\S(x) \equiv  \S ({\k x}, \k, 1)$.
To evaluate the integral in  (\ref{gvint}) consider the following
large-$x$ expansion of $g $,
%%%
 \bea
 g=&&-\fr{1+\k^2}{\k^3}\;\fr{1}{x^3}-\fr{3(1+\k^2)}{2\k^4}\;\fr{1}{x^4}
     +\fr{3(-5-3\k^2+2\k^4)}{4\k^5}\;\fr{1}{x^5} \nn\\
     &&+\fr{5(-11-5\k^2+6\k^4)}{8\k^6}\;\fr{1}{x^6}
   -\fr{3(77+21\k^2-43\k^4+10\k^6)}{16\k^7}\;\fr{1}{x^7}
 +... \  .    \label{gVexpansion}
 \eea
%%%
 Up to the
order given in (\ref{gVexpansion}) the integral yields
%%%
 \bea
 \int_{\fr{N}{\k}}^\infty dx\;g (x)
 =&& -\fr{1}{2}\left(1 + \fr{1}{\k}\right)\;\fr{\k}{N^2}
  %-\left(\fr{1}{2\k}+\fr{\k}{2}\right)\;\fr{1}{N^3}
  -\fr{1}{2}\left(1 + \fr{1}{\k}\right)\;\fr{\k}{N^3}
 \nn\\
  &&  + \fr{3}{16}\left( 2 -  \fr{3}{\k^2}  - \fr{5}{\k^4}
 \right)\;\fr{\k^3}{N^4}
  + \fr{1}{8}\left( 6 -  \fr{3}{\k^2}  - \fr{11}{\k^4}
 \right)\;\fr{\k^3}{N^5}
%  -\left(\fr{11}{8\k}+\fr{5\k}{8}
 % -\fr{3\k^3}{4}\right)\;\fr{1}{N^5}
  \nn\\
 &&
 %-\left(\fr{77}{32\k}+\fr{21\k}{32}-\fr{43\k^3}{32}
 %  +\fr{5\k^5}{16}\right)\;\fr{1}{N^6}
   +O\left(\fr{\k^5}{N^6}\right) \ .
 \eea
%%%
For $N=40000$ and $\k=50,100,200$
we find that the correction is small compared to the
numerically found value of $E_1$ (which is of order $10^{-5}$).
However, the correction grows if we increase $\k$ for fixed $N$
(e.g., consider  $\k=1000$).

%\newpage

\end{document}